\documentstyle[twoside,fleqn,espcrc2]{article}

\renewcommand{\ref}[1]{\raisebox{.6ex}{[#1]}}

\newcommand{\be}{\begin{equation}}
\newcommand{\ee}{\end{equation}}

\begin{document}

\title{ Escape of trapped electrons from a helium surface:
                a dynamical theory }

\author{ Ping Ao \\ {\ } \\  Department of Physics, FM-15,
          University of Washington, Seattle, WA 98195, USA  }

\begin{abstract}
We present a dynamical theory which incorporates
the electron-electron correlations and the effects of external magnetic fields
for an electron escaping from a helium surface.
Analytical expressions for the escape rate can be obtained in various limits.
In particular, the tunneling rate with a parallel magnetic field is presented
explicitly.
\end{abstract}



\maketitle

The system of electrons in a metastable well near a helium surface
is an ideal system to test our understanding of the escape process
from a metastable well with many body correlations due to Coulomb interaction.
A number of experiments have been performed on the escaping of electrons from a
helium surface.$^{1}$ Recent theoretical studies have mainly concentrated
on how the static correlations affect the escape rate.$^{2,3}$
In the present paper we describe a dynamical theory for the escape of an
electron from a helium surface, which accounts for the effects of
both static and dynamical correlations as well as of magnetic fields.

We consider the experimentally relevant situation in which
the lifetime of the metastable state of an electron is much longer
than its relaxation time in the metastable well
and the density of 2-d electrons is low such that the Fermi temperature
is the smallest energy scale in the problem. The escaping electrons are then
statistically independent of each other and the exchange effect
of an escaping electron with 2-d electrons can be ignored.
A separation between the escaping electron and the remaining 2-d electrons
for each escape event can be made.
Firstly we find the effective Hamiltonian which describes the escape process
in the high temperature limit.
Starting from it we use an imaginary time path integral method
to calculate the tunneling rate at low temperatures.
We shall ignore the weaker interactions between the escaping
electron and the surface waves of liquid helium and the helium vapor
atoms, which have been discussed elsewhere$^{3}$.

The key element in the present theory is to treat properly
the dynamics of the 2-d electrons
in response to the motion of the escaping electron,
which presents an induced electric force in
the equation of motion for the escaping electron.
The 2-d electron fluid is described by a set of hydrodynamical equations:
the continuity equation and the Euler's equation.
In the small density deviation
and nonrelativistic limit, we linearize the hydrodynamical equations.
We solve for the density deviation, which is determined by the motion of
plasma modes. Then using
the Poisson equation we can calculate the induced electric field$^{3}$.
At this point we find that
we are facing a problem similar to the one in the discussion of the
macroscopic quantum effect$^{4}$, where the total Hamiltonian has three
parts, a dissipative bath consisting of harmonic oscillators,
a system of interest, and the coupling between the
system and the bath. Using this analogy, we obtain the effective
Hamiltonian to describe the motion of the escaping electron:$^{5,3}$
\[
  H = \frac{1}{2m}\left[ {\bf P} - \frac{e}{c} {\bf A}_{ex} \right]^{2}
       + V_{A}(z)
       + \int_{k < k_{c} } d{\bf k} \sum_{j=1}^{2}
\]
\be
       \left[ \frac{ p_{j}^{2}({\bf k}) }{2m}
       + \frac{ m \omega_{P}^{2}(k) }{ 2 }
       \left( q_{j}({\bf k})
       - \frac{\kappa(z)\lambda_{j} }
              {m\omega_{P}^{2}(k) } \right)^{2}  \right],
\ee
with the adiabatic potential
\be
    V_{A}(z) = V_{0}(z) - \frac{1}{2} \int_{0}^{ k_{c} } dk
       \frac{\kappa^{2}(z) }{ n_{0}e^{2}(1-4\Lambda) } \; ,
\ee
the plasma dispersion relation
$ \omega_{P}^{2}(k) = \omega_{B}^{2} + n_{0}e^{2} 2\pi k
 (1-4\Lambda )/m + k_{B}T k^{2}/m $,
$k_{c} = 2\sqrt{\pi n_{0}}$,
and
$ \kappa(z)= \sqrt{ n_{0} e^{4} } (1-4\Lambda ) \exp\{-kz\} $.

Here $\omega_{B} = eB_{\bot ex} /mc $ with
$B_{\bot ex}$ the component of the external magnetic field perpendicular
to the helium surface. In the calculation the pressure $p = n k_{B}T$
for the 2-d classical electron fluid phase has been used. The effect due to
$p$ is small in the present problem.
In eq.(1) $m$ and $e$ are the mass and the charge of an electron respectively,
$c$ is the speed of light,
$\lambda_{1}=\cos({\bf k}\cdot{\bf r}) $,
$\lambda_{2}=\sin({\bf k}\cdot{\bf r}) $,
and ${\bf B}_{ex} = \nabla\times{\bf A}_{ex} $.
In eq.(2) $ V_{0}(z) = V_{w}(z) + V_{i}(z) + V_{n}(z)$,
and $V_{w}$ is the hardwall potential, $V_{w} = \infty$
for $z<0$,  $V_{w} = 0$ for $z>0$, which prevents electrons from entering the
liquid helium. The image potential is $V_{i} = - e^{2}\Lambda /z $
with $\Lambda = (\epsilon - 1 )/4(\epsilon + 1 ) $ and $\epsilon$ the
dielectric constant of liquid helium.
The potential $V_{n}$ is the total electric potential
produced by the external applied electric field (perpendicular only) and the
electric field produced by the mean density  $n_{0}$ of the 2-d electrons,
$   V_{n}(z) = - e[E_{ex} + 2\pi e (1-4\Lambda) n_{0}]z $.
The condition for 2-d electrons to escape from the surface to $z=\infty$  is
$E_{ex} + 2\pi e (1-4\Lambda)n_{0} < 0$.

If 2-d electrons completely follow the motion of the escaping electron, the
so called adiabatic limit,
their responses are described by the adiabatic potential eq.(2).
The deviation from the adiabatic response is described by the dynamics of
plasma modes, the last term in eq.(1).
The second term of eq.(2) corresponds to the correlation potential discussed
in Ref.[1]. Hence an alternative justification of its usage
in Ref.[1] is obtained here.
Based on the effective Hamiltonian, eq.(1),
the number of degree of freedom in the the calculation of escape rate is
effectively reduced from original $3N$ to 3
after integration over plasma modes,
and analytical expressions can be obtained in various limits.
The Hamiltonian may have a wider
application regime than that of the present hydrodynamical approach.
For instance, one may argue that eq.(1) can be applied to the Wigner lattice
case. This would lead to the conclusion that there is no change of escape rate
cross the melting temperature because there is no change of density
fluctuation.$^{5,3}$

For an illustration, we present the results for the situation
$B_{\| ex} \neq 0$ and $B_{\bot ex}=0$.$^{5}$
In this case, there is no influence on the thermal activation rate
because the magnetic field does not affect the barrier height,
if we ignore its small effect on the ground state energy ${\cal E}_{0}$.
For the quantum tunneling process, by a semiclassical calculation we find that
in the high field limit (The cyclotron frequency is
larger than the binding energy) the tunneling rate is zero at
zero temperature.
In the small field and low temperature limit,
the semiclassical action can be evaluated perturbatively:
\[
   S_{c} = S_{c0} + \frac{1}{2m} \left(\frac{eB_{\| ex} }{c} \right)^{2}
              \int d\tau \;  z_{c}^{2}(\tau)
\]
\be
  - \frac{1}{2m} \left(\frac{eB_{\| ex} }{c} \right)^{2} \frac{k_{B}T}{\hbar}
           \left(\int d\tau \;  z_{c}(\tau) \right)^{2} \; ,
\ee
with $S_{c0}$ the semiclassical action in the absence of the magnetic field
and the semiclassical trajectory $z_{c}(\tau)$ determined by the usual equation
$ m \dot{z}_{c}^{2}/2 = V_{A}(z_{c}) - {\cal E}_{0}$.
The tunneling rate is then $\Gamma \sim \exp\{ - S_{c}/\hbar \}$.
It should be emphasized that in the calculation leading to eq.(3)
we have taken a full consideration of the 3-d dynamics of the escaping
electron.

The author thanks A.J. Leggett for introducing this
problem and for many useful suggestions. This work was supported in part by
the MacArthur Foundation at University of Illinois,
and by the US Natural Science Foundation \#: DMR 89-16052.

\end{document}